\documentclass[superscriptaddress,notitlepage,floats,showpacs,10pt]{revtex4-1}
\usepackage{graphicx}
\usepackage{amsmath}
\usepackage{float}
\usepackage{color}

\newcommand{\re}[1]{{\color{red}#1}}
\newcounter{firstbib}

\begin{document}

\title{Paradoxical Stabilization of Forced Oscillations by  Strong Nonlinear Friction}

\author{T. Zh. Esirkepov}
\affiliation{Kansai Photon Science Institute, National Institutes for Quantum and Radiological Science and Technology (QST), 8-1-7 Umemidai, Kizugawa, Kyoto 619-0215, Japan}

\author{S. V. Bulanov}
\affiliation{Kansai Photon Science Institute, National Institutes for Quantum and Radiological Science and Technology (QST), 8-1-7 Umemidai, Kizugawa, Kyoto 619-0215, Japan}
\affiliation{A. M. Prokhorov Institute of General Physics, the Russian Academy of Sciences, Vavilov street 38, 119991 Moscow, Russia}

\begin{abstract}
In a dissipative dynamic system driven by an oscillating force,
a strong nonlinear highly oscillatory friction force
can create a quasi-steady tug, 
which is always directed opposite to the ponderomotive force
induced due to a spatial inhomogeneity of oscillations.
When the friction-induced tug exceeds the ponderomotive force,
the friction stabilizes the system oscillations
near the maxima of the  oscillation  spatial amplitude. 
\end{abstract}

\pacs{}
\date{January 27, 2017}
\maketitle

\section{Introduction}

In classical mechanics, Kapitza pendulum
or  Stephenson-Kapitza pendulum
\cite{Kapitza1, Kapitza2} is
a statically unstable inverted pendulum
whose statically unstable equilibrium position
is stabilized by small fast vertical oscillations of the pivot point.
This {\it induced stability} initially described by A.~Stephenson \cite{Stephenson}
has been first explained by P.~L.~Kapitza \cite{Kapitza1}.
To find theoretical reason of the  {\it induced stability}
Kapitza separated the pendulum motion into fast and slow oscillations
and, by averaging out fast ones,
found the effective potential 
which has minimum at the pendulum upper position, 
in contrast to a simple pendulum.
This approach created a new 
concept of {\it dynamic stabilization} in
mechanics \cite{Blekhman}.
Chelomei's pendulum provides another well known example of dynamically stabilized mechanical system 
\cite{Chelomei}.
The {\it induced stability} can be vindicated 
beyond the 
framework of the
method of fast and slow motions separation
\cite{Butikov}.

In general, a spatial inhomogeneity of an oscillating driving force
creates a quasi-steady ponderomotive force
\cite{LLM,Licht-Lieb}, 
directed against the spatial gradient of the
driving force.
In the Kapitza pendulum,
this ponderomotive force acts against gravitation and
makes the upper position stable.
A dissipation dampens oscillations
around the upper equilibrium position
thus further stabilizing it.
When no other forces present besides the oscillating driving force,
as in the case of charged particle dynamics in a standing electromagnetic wave,
the ponderomotive force always repels particles from
the maxima of the 
wave spatial amplitude \cite{Blekhman}.

Here we present a general model
of a dissipative dynamic system driven by an oscillating force,
where a strong nonlinear highly oscillatory friction
creates a quasi-steady tug,
which, quite counter-intuitively,
is always directed opposite to the ponderomotive force
and exceeds it for a sufficiently strong driving force.
This leads to a seemingly paradoxical
stabilization of the system oscillations
near the maxima of the spatial amplitude of the driving force.
It differs from the Kapitza pendulum effect
in that here the stabilization factor is a nonlinear growth of the friction 
with the driver force,
which creates a tug against the ponderomotive potential.

\section{Model}

We consider a simple one-dimensional model of a forced oscillation with a strong nonlinear friction,
given by the equation
\begin{equation}\label{eq:model}
\ddot x+\mathcal{K}(\mathcal{F}) \dot x =\mathcal{F}
, \quad \mathcal{K}=\nu \mathcal{F}^{2n}
.
 \end{equation}
Here the dot denotes differentiation with respect to time;
$n$ is a natural number.
The friction coefficient $\mathcal{K}$ is a non-negative function of the oscillating driving force $\mathcal{F}$,
\begin{equation} \label{eq:driver}
\mathcal{F}(x,t)=f_1(x)\cos \omega t+f_2(x)\sin \omega t
\, .
\end{equation}

The model is motivated by the dynamics
of a charged particle in a strong electromagnetic field,
where accelerating particles lose energy and undergo a recoil
due to their emission of electromagnetic radiation \cite{Rad-Dom}.
This causes a friction which nonlinearly grows with the
electromagnetic field strength.

Following the classical approach of Ref. \cite{LLM},
we assume that a solution to Eq. (\ref{eq:model})
can be represented as
\begin{equation} \label{eq:X-xi}
x(t)=X(t)+\xi(t)
\end{equation}
with a slowly varying function $X(t)$
and a fast oscillating small addition $\xi(t)$, $|\xi|\ll |X|$,
which has a zero time average,
\begin{equation}
 \langle \xi \rangle=(\omega/2\pi)\int_{0}^{{2\pi}/{\omega}} \xi(t) dt =0.
 \end{equation}
Correspondingly,
$ \langle x \rangle=X $,
$ \langle\dot\xi \rangle=\langle \ddot\xi \rangle =0 $, $\left< X\right>\approx X(t)$.
Substituting (\ref{eq:X-xi}) into Eq. (\ref{eq:model}) and 
expanding the functions $\mathcal{F}$ and $\mathcal{K}$ in powers of $\xi$
as
\begin{align}
\label{eq:FX}
&
\mathcal{F}\approx F+\xi \partial_X F ,\quad  F=\mathcal{F}(X,t) ;
\\
\label{eq:KX}
&
\mathcal{K}\approx K+\xi \partial_X K ,\quad  K=\mathcal{K}\big(\mathcal{F}(X,t)\big),
\end{align}
we obtain
\begin{equation} \label{eq:model-X-xi}
\ddot X+ \ddot \xi+K \dot X+K \dot \xi+
\xi \dot X \partial_X K + \xi \dot \xi \partial_X K
=F+\xi \partial_X F
\, ,
 \end{equation}
where $\partial_X$ is the partial derivative with respect to $X$
(the first argument of $\mathcal{F}$).
The time derivatives $\ddot\xi$ and $\dot\xi$
are not small, being proportional to $\omega^2$ and $\omega$, respectively.
They are assumed to be much greater than $\ddot X$ and $\dot X$.
The friction coefficient $K$ defined in Eq. (\ref{eq:KX})
is not necessarily small; it has a 
time-independent component
\begin{equation}
\kappa = \langle K \rangle >0 .
\end{equation}

In Eq. (\ref{eq:model-X-xi}),
slowly varying and fast oscillating terms should cancel out separately.
Neglecting the time derivatives of $X$,
in the zeroth order approximation 
with respect to $\xi$
we find  for the fast oscillating term
\begin{equation} \label{eq:model-xi}
\ddot \xi+K \dot \xi=F \, .
\end{equation}
Here $X$ as an argument of functions $F$ and $K$ is assumed to be constant.
The forced oscillation solution of Eq. (\ref{eq:model-xi})
can be cast in the form
\begin{gather} \label{eq:xi-sol}
\xi=e^{-\kappa t-\Delta(t)}\int_0^t e^{\kappa\tau+\Delta(\tau)} F(\tau) d\tau
,
\\
\Delta(t) = \int_0^t [ K(\tau) - \kappa ] d\tau
.
\end{gather}
The first term in the expansion 
with respect to $\Delta$ of the dependence given by Eq.  (\ref{eq:xi-sol})  
\begin{equation}\label{eq:xi}
\xi=\frac{(\kappa f_1-\omega f_2) \sin\omega t
-(\kappa f_1+\omega f_2) \cos\omega t}{\omega(\kappa^2+\omega^2)}
\, 
\end{equation}
approximates the first harmonic of the solution.

Averaging Eq. (\ref{eq:model-X-xi}) over time
and  taking into account that $\left< F(X,t)\right>\approx 0$ for nearly constant $X(t)$, we obtain
\begin{equation}
\label{eq:model-X}
\ddot X+(\kappa+\langle\xi\partial_X K\rangle) \dot X=
\langle \xi \partial_X F \rangle -
\langle \xi \dot \xi \partial_X K\rangle
\, .
\end{equation}
Using here the expression for $\xi$, given by Eq.  (\ref{eq:xi}),
and the definitions for $F$ and $K$ formulated above,
we obtain
the equation for the slowly varying function $X(t)$
and the average friction coefficient:
\begin{align}
\label{eq:X}
\ddot X&+\kappa \dot X=
-\frac{\partial _X (f_1^2+f_2^2)}
         {4(\kappa^2+\omega^2)}
+
\frac{n^2\kappa^2 \partial _X (f_1^2+f_2^2)}
        {(n+1)(\kappa^2+\omega^2)^2}
+
\frac{\kappa (f_2 \partial_X f_1 - f_1 \partial_X f_2)}
        {2\omega(\kappa^2+\omega^2)}
\left[
\frac{2n}{n+1}\left(
\frac{\kappa^2-\omega^2}{\kappa^2+\omega^2}\right)
-1
\right]
,
\\&
\textstyle
\kappa=
2^{-2n}\binom{2n}{n}
\nu
(f_1^2+f_2^2)^n
,
\label{eq:kappa}
\end{align}
where $\binom{2n}{n}$ is the binomial coefficient.
The first term on the r.h.s. of Eq. (\ref{eq:X}) 
corresponds to the classical ponderomotive force \cite{LLM}, 
modified due to the friction,
the last terms represent the tug induced by the friction.
In the case of $f_2=0$, 
the third term in the r.h.s. of Eq. (\ref{eq:X}) vanishes.
Then the tug becomes always directed opposite to the ponderomotive force.
It can even exceed the latter in magnitude,
when
\begin{equation}
\nu > 
\frac{2^{2n} (n!)^2 (n+1)^{1/2}}
        {(2n)! (4n^2-n-1)^{1/2}}
\frac{\omega}{f_1^{2n}}
\, .
\end{equation}
In this case, trajectories of the system described by the model (\ref{eq:model})
drift to the local maximum of the driving force spatial amplitude.
If the system were non-dissipative,
the oscillations near that maximum would be destabilized
by the ponderomotive force, so that
the corresponding trajectories would drift 
{\it against} the spatial gradient of the driving force
and would eventually escape to regions of a lower spatial amplitude 
of the driving force \cite{Blekhman}.
Sufficiently strong friction makes the trajectories to drift 
{\it along} the spatial gradient of the driving force,
provided that these trajectories have already got to a region with sufficiently high driving force.
This causes a seemingly paradoxical
stabilization of the oscillations
near the local maximum of the driving force spatial amplitude.

\section{Numerical simulations}

\begin{figure}[H]
\begin{center}
  \includegraphics[width=0.7\columnwidth]{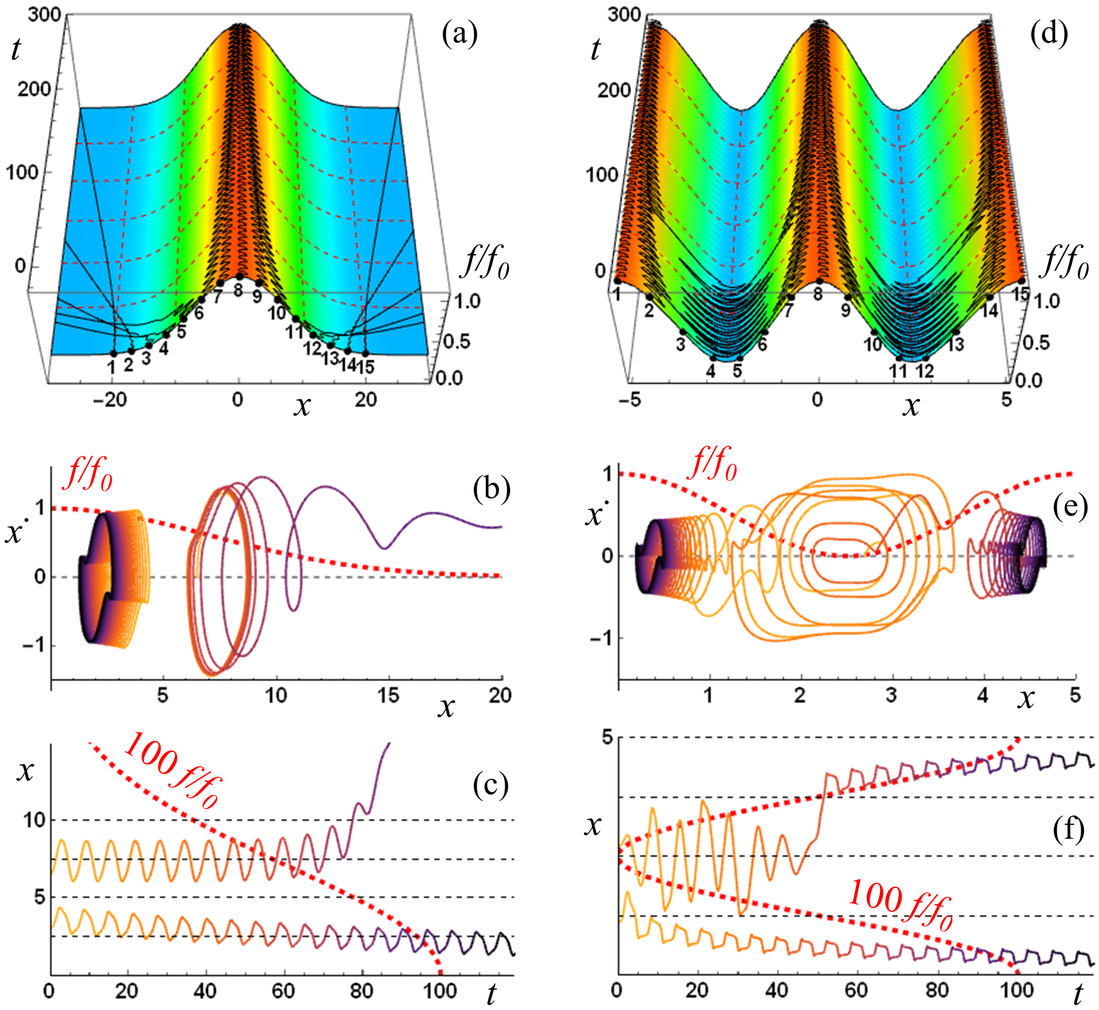}
\end{center}
\caption{Numerical solutions of Eq. (\ref{eq:model}) in the case of 
the bell-shaped (a,b,c) and spatially periodic (d,e,f) driving force.
%
(a,d) Trajectories numbered from 1 to 15,  started from various locations with
$\dot x_{t=0}=0$, plotted over the driving force spatial amplitude normalized to the maximum.
(b,e) Phase portrait for two representative trajectories in the $(x,\dot x)$ plane.
(c,f) The same trajectories in terms of $x(t)$; 
the dashed curve is for the driving force spatial amplitude (rescaled).
\label{fig:num}
}
\end{figure}

The effect of the stabilization of oscillations due to strong friction
is further demonstrated by the numerical integration of the model equation (\ref{eq:model})
with the friction coefficient taken in the form $\mathcal{K}=\nu \mathcal{F}(x,t)^4$
in two cases.
In the first case, shown in Fig. \ref{fig:num} (a,b,c),
the driving force amplitude is bell-shaped,
$\mathcal{F}(x,t)=f_0\exp(-(x/l_0)^2) \cos \omega t$, 
with the width, oscillation frequency, amplitude, and  
friction factor equal to
$l_0=10$, $\omega=1$, $f_0=3$, and $\nu=0.2$, 
respectively.
In the second case, Fig. \ref{fig:num} (d,e,f),
the driving force is spatially periodic,
$\mathcal{F}(x,t)=f_0\cos^2(2\pi x/l_0) \cos \omega t$,
$l_0=10$, $\omega=1$, $f_0=8$, and $\nu=0.25$.

In the case of bell-shaped driving force shown in Fig. \ref{fig:num} (a,b,c),
the trajectories, starting at $t=0$ from locations 
where the driving force spatial amplitude is relatively weak,
exhibit several oscillations and then escape
being pushed away by the ponderomotive force.
In the region of a high spatial amplitude of the driving force,
the friction-induced tug overcomes the ponderomotive force,
therefore the trajectories started from this region drift towards the
maximum of the driving force spatial amplitude.
As they get closer to that maximum,
their drift becomes slower and their oscillation amplitude decreases.

In the case of spatially periodic driving force (see Fig. \ref{fig:num} (d,e,f)),
the trajectories, initially oscillating near the 
minima of the driving force spatial amplitude
and having enough large oscillation amplitude,
reach the region of higher driving  force,
where the trajectories are caught by the friction-induced tug.
Eventually all such trajectories are reduced to 
oscillations near the maxima of the driving force spatial amplitude.

In both cases the drift is slowing down
near the maxima of the driving force spatial amplitude 
in agreement with Eq. (\ref{eq:X}),
because the gradient $\partial_X f$
vanishes at the maximum of $f$.

%

\section{Limit cycle}

On a trajectory asymptotically turning into periodic oscillations
seen in Fig. \ref{fig:num} (b,c,e,f),
the driving force is almost constant,
which simplifies theoretical consideration allowing 
more detail description of the driven oscillations.
This case is described by the approximation of 
$\mathcal{F}(x,t)=f_0 \cos \omega t$. 
We change variables to $(\tau,y)$
and introduce the friction parameter $\sigma$:
\begin{gather}
t = \tau/\omega, \quad x(t) = (f_0/\omega^2)y(\tau)  ;
\label{eq:model:newvars}
\\
\sigma=\nu f_0^{2n}/\omega .
\label{eq:model:sigma}
\end{gather}
Then Eq. (\ref{eq:model}) becomes
\begin{equation}
y''(\tau) + \sigma \cos^{2n}(\tau) y'(\tau) = \cos \tau
\, .
\label{eq:model:fconst}
\end{equation}
The general solution can be represented as
\begin{align}
&
y'(\tau)=
e^{-\sigma S_n(\tau)}
[ y'(0) - {Y}_{n}(0) ] +  {Y}_{n}(\tau)
\, ,
 \label{eq:model:fconst:sol}
\\
&
{Y}_{n} (\tau)  =
\frac{e^{-\sigma S_n(\tau)}}{e^{\pi\alpha\sigma}+1}
\!\!\!
\int\limits_{-\pi/2}^{\pi/2}
\!\!\!\!
e^{\sigma S_n(\zeta+\tau+\frac{\pi}{2})}
\sin(\zeta \! + \! \tau)d\zeta
,
\label{eq:model:fconst:LC}
\\
&
\textstyle
S_n(\tau) = 
\alpha \tau+
\sum\limits_{m=1}^{n}
\binom{2n}{n+m} \frac{\sin(2m\tau)}{2^{2n}m}
,
\ \ 
\alpha=2^{-2n}\binom{2n}{n}
.
\end{align}
As one can see, any solution at $\tau\rightarrow\infty$ 
tends to the limit cycle described by the function ${Y}_{n}$.

The amplitude of the limit cycle in terms of the derivative $y'$,
${A}_{n}=\max|{Y}_{n}(\tau+\pi)-{Y}_{n}(\tau)|$, is
\begin{equation}
{A}_{n}=
\frac{1}{\cosh(\frac{\pi\alpha\sigma}{2})}
\!\!\!
\int\limits_{-\pi/2}^{\pi/2}
\!\!\!\!
e^{\sigma S_n(\tau)}\cos(\tau)d\tau
.
\end{equation}
For large $\sigma$, it decreases  as a negative power of $\sigma$,
e.g., ${A}_1 \approx 1.88\sigma^{-2/3}$.
Correspondingly, the oscillation amplitude decreases
when the trajectory drifts towards the driving force maximum,
as seen in Fig. \ref{fig:num}.

\begin{figure}[H]
\begin{center}
\includegraphics[width=0.5\columnwidth]{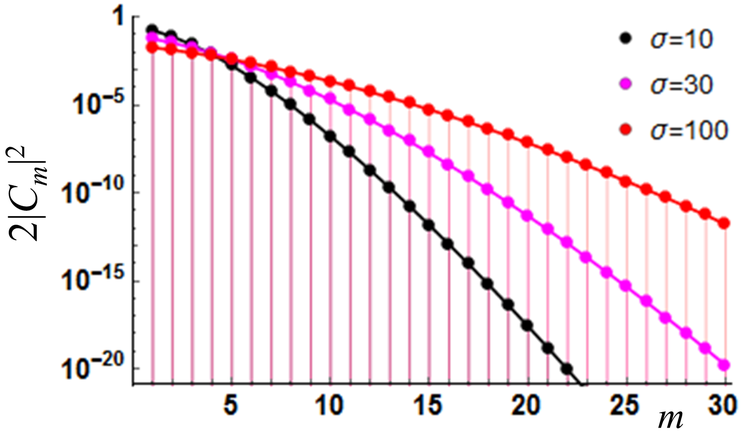}
\end{center}
\caption{Spectral density of the function ${Y}_1$,
Eq. (\ref{eq:LCF}), corresponding to $\dot x$ of Eq. (\ref{eq:model})
 in the limit cycle,
for several values of the friction parameter
$\sigma=\nu f_0^{2}/\omega$.
\label{fig:spectrum}
}
\end{figure}

In the case of $n=1$,
the function describing the limit cycle, Eq. (\ref{eq:model:fconst:LC}),
can be represented as a Fourier series
in terms of odd harmonics of the driving force
\begin{equation}
\textstyle
{Y}_1(\tau)\!=
\!\!
\sum\limits_{m=1}^{\infty}
\!
\left[
C_{m}e^{i(2m-1)\tau}+C_{m}^*e^{-i(2m-1)\tau}
\right]
\!\!
\, ,
\label{eq:LCF}
\end{equation}
where the asterix denotes complex conjugation.
The sequence $C_1$, $C_m$, $m\ge2$,
representing the frequency spectrum of ${Y}_1$,
is expressed in terms of modified Bessel functions
of the first kind, $I_k$:
\begin{gather}
C_1=
\!\!\!
\sum\limits_{k=-\infty}^{\infty}
\!\!\!
\frac{(-1)^{k+1}}{4k+2-i\sigma}
\left[
i I_k^2\left(\frac{\sigma}{4}\right) + I_{k}\left(\frac{\sigma}{4}\right)I_{k+1}\left(\frac{\sigma}{4}\right)
\right]
\, ,
\label{eq:C1}
\\
C_m =
i^m
\frac{I_{m-\frac{1}{2}-\frac{i\sigma}{4}}\left(\frac{\sigma}{4}\right)}%
       {I_{\frac{3}{2}-\frac{i\sigma}{4}}\left(\frac{\sigma}{4}\right)}
\left[
\left(2+\frac{4i}{\sigma}\right)C_1 +C_1^* - \frac{2}{\sigma}
\right]\!\!
.
\label{eq:Cm}
\end{gather}
The amplitude of the first harmonic, $C_1$, as a function of $\sigma$ can be approximated by
\begin{equation}
C_1(\sigma)\approx 
\frac{\sigma/2}{4+\sigma^{2-\delta}}
-
i\left[
\frac{\sigma/3-2}{4+\sigma^{2-\delta}}
+
\frac{16+3\sigma^2}{(4+\sigma^{2-\delta})^2}
\right]
\end{equation}
with $\delta\approx 0.098$.
The spectral density of the 
function ${Y}_1(\tau)$,
Eq. (\ref{eq:LCF}),
is $|2C_{m}|^2$;
it is shown in
Fig. \ref{fig:spectrum}.
As one can see, 
the frequency spectrum contains high order harmonics according to Eqs. (\ref{eq:LCF}-\ref{eq:Cm}).

\section{Conclusion}

In conclusion, in contrast to known {\it dynamical destabilization} 
under the action of dissipation known as dissipation-induced instabilities 
(see review article \cite{Destabilization} and references therein) we show that
a {\it strong nonlinear friction} can cause
a seemingly paradoxical stabilization of forced oscillations
near the maxima of the driving force spatial amplitude.
In particular, such a friction
occurs in the dynamics of charged particles
in ultra-strong electromagnetic fields
due to radiation reaction \cite{Rad-Dom}.
The threshold for the described stabilization of forced oscillations
corresponds to the criterion of the importance of the radiation reaction effects
\cite{Survey}.

In a standing electromagnetic waves,
which can be formed in multiple high power laser configurations \cite{Multi-las},
the stabilization due to a strong radiation reaction
is manifested in
an anomalous electron bunching near the electric field maxima \cite{Anom-bunch}.
As shown in Refs. \cite{Anom-bunch, Fedotov,Attr, Jirka, Kirk},
electrons  can be captured for many laser periods due to
radiation friction impeding the ponderomotive force.
When radiation reaction dominates,
the electron motion in a standing wave 
evolves to limit cycles and strange attractors \cite{Attr, Jirka}.
In the case of a circularly polarized standing wave,
analytical expressions exist for the limit cycles
near the electric field maxima \cite{Kirk}.
A collision of multiple ultra-intense electromagnetic waves
creates structurally determinate patterns 
in the electron phase space \cite{Survey,Vranic,Gong}
due to a counterplay of the ponderomotive force
and the friction-induced tug.

Although in the present work we have been motivated 
by the intention to build up the theory of the radiative electron dynamics  
in the field of extremely high intensity lasers, we believe that the formulated above 
concept of dissipative stabilization of nonlinear dynamic systems will be useful for 
applications well beyond the framework of the 
laser-matter interaction physics \cite{MTB,Marklund},
remembering a saying of William Thomson (Lord Kelvin) 
``I never satisfy myself until I can make a mechanical model of a thing''
\cite{Kelvin}.
%


\numberwithin{equation}{subsection}

\section*{Appendix}

Here we present mathematical derivations for some formulae shown above.
Notations for variables are the same as in the main text.
The equation numbering is preserved for those equations which appear in the main text;
auxiliary formulae are numbered within sections.

\subsection{Equations (14) and (15)}
In this section we derive Eqs. \eqref{eq:X} and \eqref{eq:kappa} from Eq. \eqref{eq:model-X}.
We use the following formulae explicitly written or assumed in the main text:
\begin{gather}
F = f_1 \cos(\omega t) + f_2 \sin(\omega t),
\\
K = \nu \left[ f_1 \cos(\omega t) + f_2 \sin(\omega t) \right]^{2n},
\\
\tag{{12}}
\xi=\frac{(\kappa f_1-\omega f_2) \sin\omega t
-(\kappa f_1+\omega f_2) \cos\omega t}{\omega(\kappa^2+\omega^2)}
\, ,
\\
\tag{{13}}
\ddot X+(\kappa+\langle\xi\partial_X K\rangle) \dot X=
\langle \xi \partial_X F \rangle -
\langle \xi \dot \xi \partial_X K\rangle
\, .
\end{gather}
The time-averaged friction coefficient, defined in Eq. (8) in the main text, is
\begin{gather}
\begin{split}
&\kappa=\langle K\rangle=
\frac{\omega}{2\pi}\int_{0}^{2\pi/\omega}
\nu \left[ f_1 \cos(\omega t) + f_2 \sin(\omega t) \right]^{2n} dt=
\frac{1}{2\pi}\int_{0}^{2\pi}
\nu \left[ \frac{f_1-i f_2}{2}e^{i \tau} + \frac{f_1+i f_2}{2}e^{-i \tau}  \right]^{2n} d\tau=
\\
&=\frac{\nu}{2\pi}\sum\limits_{m=0}^{2n}
\binom{2n}{m}
\left(\frac{f_1+i f_2}{2}\right)^{m}
\left(\frac{f_1-i f_2}{2}\right)^{2n-m}
\re{
\int_{0}^{2\pi}
e^{2(n-m)i\tau}
d\tau
}=
\frac{\nu}{2\pi}
\binom{2n}{n}
\left(\frac{f_1+i f_2}{2}\right)^{n}
\left(\frac{f_1-i f_2}{2}\right)^{n}
\re{2\pi}
.
\end{split}
\end{gather}
Here $\binom{2n}{m}=(2n)!/[m!(2n-m)!]$ is the binomial coefficient.
The last integral (marked with red) is zero except the case $m=n$,
for which it equals $2\pi$ (therefore the sum contains the only nonzero term,
for the index $m=n$).
In this way we obtain Eq. \eqref{eq:kappa} from the main text:
\begin{align}
\tag{{15}}
\kappa=
2^{-2n}\binom{2n}{n}
\nu
(f_1^2+f_2^2)^n
.
\end{align}

In Eq. \eqref{eq:model-X},
the term $\langle\xi\partial_X K\rangle$ 
is zero because
it contains only odd harmonics,
\begin{equation}
\langle\xi\partial_X K\rangle = 
\frac{\omega}{2\pi}\int_{0}^{2\pi/\omega}
\xi(t) \times
2n \nu \left[ f_1 \cos(\omega t) + f_2 \sin(\omega_t) \right]^{2n-1}
 \left[ \partial_X f_1 \cos(\omega t) + \partial_X f_2 \sin(\omega t) \right]
dt
=0.
\end{equation}
The next averaged term is obtained by a simple integration
\begin{gather}
\begin{split}
\langle \xi \partial_X F \rangle = 
\frac{\omega}{2\pi}\int_{0}^{2\pi/\omega}
\frac{(\kappa f_1-\omega f_2) \sin\omega t-(\kappa f_1+\omega f_2) \cos\omega t}{\omega(\kappa^2+\omega^2)}
\left[ f_1 \cos(\omega t) + f_2 \sin(\omega_t) \right]
dt=
\\
=
-\frac{f_1 \partial_X f_1 + f_2 \partial_X f_2}{2(\kappa^2+\omega^2)}
-\frac{\kappa(f_2 \partial_X f_1 - f_1 \partial_X f_2)}{2\omega(\kappa^2+\omega^2)}
.
\end{split}
\end{gather}
The last averaged terms is
\begin{gather}
\begin{split}
-\langle \xi \dot \xi \partial_X K\rangle &= 
-\frac{\omega}{2\pi}\int_{0}^{2\pi/\omega}
\xi(t)\dot\xi(t)\times
2n \nu \left[ f_1 \cos(\omega t) + f_2 \sin(\omega_t) \right]^{2n-1}
 \left[ \partial_X f_1 \cos(\omega t) + \partial_X f_2 \sin(\omega t) \right]
dt=
\\
&=
-
\frac{2n\nu}{8\omega(\kappa^2+\omega^2)^2}
\sum\limits_{m=0}^{2n-1}
\binom{2n-1}{m}
\left(\frac{f_1+i f_2}{2}\right)^{m}
\left(\frac{f_1-i f_2}{2}\right)^{2n-1-m}
\times
\\
&
\times
\bigg\{
(\kappa - i\omega)^2(f_1-i f_2)^2
	\Big[
	(\partial_X f_2 -i\partial_X f_1) \delta_{m,n}
	-i(\partial_X f_1 -i\partial_X f_2) \delta_{m,n+1}
	\Big]
+
\bigg.
\\
&
\bigg.
+
(\kappa + i\omega)^2(f_1+i f_2)^2
\times
	\Big[
	(\partial_X f_2 + i\partial_X f_1) \delta_{m,n-1}
	+i(\partial_X f_1 + i\partial_X f_2) \delta_{m,n-2}
	\Big]
\bigg\}
=
\\
&=
\frac{n \kappa}{(n+1)\omega(\kappa^2+\omega^2)^2}
\bigg[
2n\kappa\omega(f_1 \partial_X f_1 + f_2 \partial_X f_2)
+(\kappa^2-\omega^2)(f_2 \partial_X f_1 - f_1 \partial_X f_2)
\bigg].
\end{split}
\end{gather}
Here $\delta_{m,n}$ is the Kronecker delta;
it equals one for $m=n$ and zero otherwise.
In the last line $\nu$ is expressed in terms of $\kappa$ using Eq. \eqref{eq:kappa}.
Combining the results  into Eq. \eqref{eq:model-X}
we obtain Eq. \eqref{eq:X} from the main text:
\begin{align}
\tag{{14}}
\ddot X+\kappa \dot X=
-\frac{\partial _X (f_1^2+f_2^2)}
         {4(\kappa^2+\omega^2)}
+
\frac{n^2\kappa^2 \partial _X (f_1^2+f_2^2)}
        {(n+1)(\kappa^2+\omega^2)^2}
+
\frac{\kappa (f_2 \partial_X f_1 - f_1 \partial_X f_2)}
        {2\omega(\kappa^2+\omega^2)}
\left[
\frac{2n}{n+1}\left(
\frac{\kappa^2-\omega^2}{\kappa^2+\omega^2}\right)
-1
\right]
.
\end{align}

\subsection{Equations (20-22)}

In this section we derive Eqs. \eqref{eq:model:fconst:sol},
\eqref{eq:model:fconst:LC} and \eqref{eq:Sn},
representing the solution of Eq. \eqref{eq:model:fconst}:
\begin{equation}\tag{{19}}
y''(\tau) + \sigma \cos^{2n}(\tau) y'(\tau) = \cos \tau
\, .
\end{equation}
We seek the solution for the first derivative $y'$ in the form
\begin{equation}\label{eq:anz}
y'(\tau) = M(\tau)\exp\left(-\sigma\int_0^\tau \cos^{2n}(\eta)d\eta\right)
.
\end{equation}
Substituting this ansatz into Eq. \eqref{eq:model:fconst},
we obtain 
\begin{equation}
M'(\tau) = \exp\left(\sigma\int_0^\tau \cos^{2n}(\eta)d\eta\right)\cos(\tau)
.
\end{equation}
Integrating this equation for $M$ and substituting the result into Eq. \eqref{eq:anz},
we obtain a general solution of Eq. \eqref{eq:model:fconst}:
\begin{equation}\label{eq:gen}
y'(\tau) = \exp\left(-\sigma\int_0^\tau \cos^{2n}(\eta)d\eta\right)
\left[y'(0) +
\int_0^\tau
\exp\left(\sigma\int_0^\zeta \cos^{2n}(\eta)d\eta\right)
\cos(\zeta)
d\zeta
\right]
.
\end{equation}
The term in the exponent can be expanded as
\begin{gather}
\begin{split}
S_n(\tau) 
&= \int_0^\tau \cos^{2n}(\eta)d\eta
= \int_0^\tau \left(\frac{e^{i\eta}+e^{-i\eta}}{2}\right)^{2n}d\eta
=\int_0^\tau \sum\limits_{m=0}^{2n} \frac{1}{2^{2n}} \binom{2n}{m}
e^{i (2n-m)\eta}e^{-i m\eta}d\eta
=
\\
&=\frac{1}{2^{2n}}\sum\limits_{m=0}^{2n} \binom{2n}{m}
\int_0^\tau e^{2i (n-m)\eta}d\eta
= \frac{1}{2^{2n}}\binom{2n}{n}\tau +
\frac{1}{2^{2n}}\sum\limits_{\substack{m=0\\ m\not=n}}^{2n} \binom{2n}{m}
\frac{e^{2i (n-m)\tau} - 1}{2i(n-m)}
=
\\
&=\frac{1}{2^{2n}}\binom{2n}{n}\tau +
\frac{1}{2^{2n}}\sum\limits_{\substack{m=0\\ \re{m\rightarrow n-m'}}}^{n-1} \binom{2n}{m}
\frac{e^{2i (n-m)\tau} - 1}{2i(n-m)}
+
\frac{1}{2^{2n}}\sum\limits_{\substack{m=n+1\\ \re{m\rightarrow n+m'}}}^{2n} \binom{2n}{m}
\frac{e^{2i (n-m)\tau} - 1}{2i(n-m)}
=
\\
&=\frac{1}{2^{2n}}\binom{2n}{n}\tau +
\frac{1}{2^{2n}}\sum\limits_{m'=1}^{n} \binom{2n}{n-m'}
\frac{e^{2i m' \tau} - 1}{2i m'}
+
\frac{1}{2^{2n}}\sum\limits_{m'=1}^{n} \binom{2n}{n+m'}
\frac{e^{- 2i m'\tau} - 1}{-2i m'}
=
\\
&=\frac{1}{2^{2n}}\binom{2n}{n}\tau +
\frac{1}{2^{2n}}\sum\limits_{m=1}^{n} \binom{2n}{n+m}
\frac{e^{2i m \tau} - e^{-2i m \tau}}{2i m}
\\
&=\frac{1}{2^{2n}}\binom{2n}{n}\tau +
\frac{1}{2^{2n}}\sum\limits_{m=1}^{n} \binom{2n}{n+m}
\frac{\sin(2m \tau)}{m}
.
\end{split}
\end{gather}
Here we rearranged the sums changing to new indices (marked with red)
and used the identity
$\binom{2n}{n-m}=\binom{2n}{n+m}$ for ${n\ge m\ge 0}$.
In this way we obtain Eq. \eqref{eq:Sn} in the main text:
\begin{equation}
S_n(\tau) = 
\alpha \tau+
\sum\limits_{m=1}^{n}
\binom{2n}{n+m} \frac{\sin(2m\tau)}{2^{2n}m}
,
\ \ 
\alpha=2^{-2n}\binom{2n}{n}
.
\label{eq:Sn}
\tag{{22}}
\end{equation}
Below we use the following properties of the function $S_n$:
\begin{equation}
S_n(-\tau)=S_n(\tau), \quad
S_n(\tau+\pi)=S_n(\tau)+\pi\alpha, \quad
S_n(\pi/2)=\pi\alpha/2.
\end{equation}

The limit cycle is a periodic solution $Y_n$ of Eq. \eqref{eq:model:fconst},
therefore for any $\tau$
\begin{equation}
Y_n(\tau) = Y_n(\tau+2\pi).
\end{equation}
Using \eqref{eq:gen} we rewrite this expression as
\begin{gather}
\begin{split}
Y_n(\tau)
&=
e^{-\sigma S_n(\tau+2\pi)}
\left[y'(0) +
\int\limits_0^{\tau+2\pi}
e^{\sigma S_n(\zeta)} \cos(\zeta)
d\zeta
\right]
=
\\
&=
e^{-2\pi\alpha\sigma} \re{e^{-\sigma S_n(\tau)}}
\left[
\re{y'(0) +
\int\limits_0^{\tau}
e^{\sigma S_n(\zeta)} \cos(\zeta)
d\zeta
}
+
\int\limits_{\tau}^{\tau+2\pi}
e^{\sigma S_n(\zeta)} \cos(\zeta)
d\zeta
\right]
=
\\
&=
e^{-2\pi\alpha\sigma}
\left[
\re{Y_n(\tau)} +
e^{-\sigma S_n(\tau)}
\int\limits_{\tau}^{\tau+2\pi}
e^{\sigma S_n(\zeta)} \cos(\zeta)
d\zeta
\right]
,
\end{split}
\end{gather}
which gives the formula for $Y_n(\tau)$
\begin{gather}
\begin{split}
Y_n(\tau) 
&=
\frac{e^{-\sigma S_n(\tau)}}{e^{2\pi\alpha\sigma}-1}
\int\limits_{\tau}^{\tau+2\pi}
e^{\sigma S_n(\zeta)} \cos(\zeta)
d\zeta
=
\frac{e^{-\sigma S_n(\tau)}}{e^{2\pi\alpha\sigma}-1}
\int\limits_{0}^{2\pi}
e^{\sigma S_n(\zeta+\tau)} \cos(\zeta+\tau)
d\zeta=
\\
&=\frac{e^{-\sigma S_n(\tau)}}{e^{2\pi\alpha\sigma}-1}
\left[
\int\limits_{0}^{\pi}
e^{\sigma S_n(\zeta+\tau)} \cos(\zeta+\tau)
d\zeta
+\int\limits_{\pi}^{2\pi}
e^{\sigma S_n(\zeta+\tau)} \cos(\zeta+\tau)
d\zeta
\right]=
\\
&=
\frac{e^{-\sigma S_n(\tau)}}{e^{2\pi\alpha\sigma}-1}
\left[
\int\limits_{0}^{\pi}
e^{\sigma S_n(\zeta+\tau)} \cos(\zeta+\tau)
d\zeta
-\int\limits_{0}^{\pi}
e^{\sigma S_n(\zeta+\tau)+\pi\alpha\sigma} \cos(\zeta+\tau)
d\zeta
\right]=
\\
&=
\frac{e^{-\sigma S_n(\tau)}}{e^{2\pi\alpha\sigma}-1}
(1-e^{\pi\alpha\sigma})
\int\limits_{0}^{\pi}
e^{\sigma S_n(\zeta+\tau)} \cos(\zeta+\tau)
d\zeta
=
\frac{e^{-\sigma S_n(\tau)}}{e^{\pi\alpha\sigma}+1}
\int\limits_{-\pi/2}^{\pi/2}
e^{\sigma S_n(\zeta+\tau+\pi/2)} \sin(\zeta+\tau)
d\zeta
.
\end{split}
\end{gather}
In this way we obtain Eq. \eqref{eq:model:fconst:LC} in the main text:
\begin{equation}
{Y}_{n} (\tau)  =
\frac{e^{-\sigma S_n(\tau)}}{e^{\pi\alpha\sigma}+1}
\!\!\!
\int\limits_{-\pi/2}^{\pi/2}
\!\!\!\!
e^{\sigma S_n(\eta+\tau+{\pi}/{2})}
\sin(\eta \! + \! \tau)d\eta
.
\tag{{21}}
\end{equation}
Accepting $Y_n$ as a particular solution of Eq. \eqref{eq:model:fconst}
and rewriting Eq. \eqref{eq:gen},
we obtain another form of the general solution of Eq. \eqref{eq:model:fconst},
namely Eq. \eqref{eq:model:fconst:sol}:
\begin{equation}
y'(\tau)=
e^{-\sigma S_n(\tau)}
[ y'(0) - {Y}_{n}(0) ] +  {Y}_{n}(\tau)
\, .
\tag{{20}}
\end{equation}

\subsection{Equation (23)}
In this section we derive Eq. \eqref{eq:An}.
The maximum difference $|{Y}_{n}(\tau+\pi)-{Y}_{n}(\tau)|$
is reached for $\tau=-\pi/2$,
therefore
\begin{gather}
\begin{split}
{A}_{n}
&={Y}_{n}(\pi/2)-{Y}_{n}(-\pi/2)=
\\
&=
\frac{e^{-\sigma S_n(\pi/2)}}{e^{\pi\alpha\sigma}+1}
\!\!\!
\int\limits_{-\pi/2}^{\pi/2}
\!\!\!\!
e^{\sigma S_n(\eta+{\pi})}
\sin(\eta \! + \! \pi/2)d\eta
-
\frac{e^{-\sigma S_n(-\pi/2)}}{e^{\pi\alpha\sigma}+1}
\!\!\!
\int\limits_{-\pi/2}^{\pi/2}
\!\!\!\!
e^{\sigma S_n(\eta)}
\sin(\eta \! - \! \pi/2)d\eta
=
\\
&=
\frac{e^{-\pi\alpha\sigma/2}}{e^{\pi\alpha\sigma}+1}
\!\!\!
\int\limits_{-\pi/2}^{\pi/2}
\!\!\!\!
e^{\sigma S_n(\eta)+\pi\alpha\sigma}
\cos(\eta)d\eta
+
\frac{e^{\pi\alpha\sigma/2}}{e^{\pi\alpha\sigma}+1}
\!\!\!
\int\limits_{-\pi/2}^{\pi/2}
\!\!\!\!
e^{\sigma S_n(\eta)}
\cos(\eta)d\eta
=
\\
&=\frac{2e^{\pi\alpha\sigma/2}}{e^{\pi\alpha\sigma}+1}
\!\!\!
\int\limits_{-\pi/2}^{\pi/2}
\!\!\!\!
e^{\sigma S_n(\eta)}
\cos(\eta)d\eta
=\frac{1}{\cosh(\frac{\pi\alpha\sigma}{2})}
\!\!\!
\int\limits_{-\pi/2}^{\pi/2}
\!\!\!\!
e^{\sigma S_n(\eta)}\cos(\eta)d\eta
.
\end{split}
\end{gather}
Thus we obtain Eq. \eqref{eq:An} in the main text:
\begin{equation}\label{eq:An}
\tag{{23}}
{A}_{n}=
\frac{1}{\cosh(\frac{\pi\alpha\sigma}{2})}
\!\!\!
\int\limits_{-\pi/2}^{\pi/2}
\!\!\!\!
e^{\sigma S_n(\tau)}\cos(\tau)d\tau
.
\end{equation}

\subsection{Equations (25) and (26)}
In this section we derive Eqs. \eqref{eq:C1} and \eqref{eq:Cm}
for the case of $n=1$.
The function $Y_1$ is a periodic solution
of the equation
\begin{equation}
Y'_1(\tau) + \sigma \cos^{2}(\tau) Y_1(\tau) = \cos \tau
\, .
\label{eq:n2}
\end{equation}
It is represented as a Fourier series, which
obviously should contain only odd hamonics of the driver:
\begin{equation}
{Y}_1(\tau)\!=
\!\!
\sum\limits_{m=1}^{\infty}
\!
\left[
C_{m}e^{i(2m-1)\tau}+C_{m}^*e^{-i(2m-1)\tau}
\right]
\!\!
\, ,
\tag{{24}}
\end{equation}
where symbol ``*'' denotes complex conjugation.
Substituting Eq. \eqref{eq:LCF}
into Eq. \eqref{eq:n2}
we obtain
\begin{equation}
\begin{split}
&\sum\limits_{m=1}^{\infty}
\left[
i(2m-1) C_{m}e^{i(2m-1)\tau} - i(2m-1) C_{m}^*e^{-i(2m-1)\tau}
\right]
+
\\
&+
\sigma\left(\frac{e^{2i\tau}}{4}+\frac{1}{2}+\frac{e^{-2i\tau}}{4}\right)
\sum\limits_{m=1}^{\infty}
\left[
C_{m}e^{i(2m-1)\tau} + C_{m}^*e^{-i(2m-1)\tau}
\right]
=\frac{e^{i\tau}}{2}+\frac{e^{-i\tau}}{2}
.
\end{split}
\end{equation}
Rearranging the sums and collecting the Fourier coefficients of the terms corresponding to the same harmonics, we easily find the following recursive relations:
\begin{gather}
C_2=-\left[
\left(2+\frac{4i}{\sigma}\right)C_1 +C_1^* - \frac{2}{\sigma}
\right] , \label{eq:C2r}
\\
i^{-m} C_m = \frac{\sigma/4}{2(m-1/2-i\sigma/4)}
\left(
i^{-(m-1)} C_{m-1} - i^{-(m+1)} C_{m+1}
\right). \label{eq:Cmr}
\end{gather}
The last expression is the same as a recurrence identity for the 
modified Bessel function of the first kind \cite{math},
\begin{equation}
I_\mu(z) = \frac{z}{2\mu}(I_{\mu-1}(z)-I_{\mu+1}(z))
\end{equation}
for $z=\sigma/4$ and $\mu=m-1/2-i\sigma/4$,
and $I_\mu(z) \propto i^{-m} C_m$.
The general solution of Eq. \eqref{eq:Cmr} is
a linear combination of the modified Bessel functions of
the first and second kinds, where 
one should cancel out an unbounded term:
\begin{equation}
C_m = U i^m I_{m-1/2-i\sigma/4}\left(\frac{\sigma}{4}\right).
\end{equation}
The coefficient of proportionality, $U$, is determined using the relation Eq. \eqref{eq:C2r}:
\begin{equation}
C_2 = -U I_{3/2-i\sigma/4}\left(\frac{\sigma}{4}\right)
=-\left[
\left(2+\frac{4i}{\sigma}\right)C_1 +C_1^* - \frac{2}{\sigma}
\right]
.
\end{equation}
As a result we obtain Eq. \eqref{eq:Cm} of the main text:
\begin{equation}
C_m =
i^m
\frac{I_{m-\frac{1}{2}-\frac{i\sigma}{4}}\left(\frac{\sigma}{4}\right)}%
       {I_{\frac{3}{2}-\frac{i\sigma}{4}}\left(\frac{\sigma}{4}\right)}
\left[
\left(2+\frac{4i}{\sigma}\right)C_1 +C_1^* - \frac{2}{\sigma}
\right]\!\!
.
\tag{{26}}
\end{equation}

The formula Eq. \eqref{eq:C1} for the coefficient $C_1$
can be obtained using series representation for the function $Y_1$.
For $n=1$, we have $\alpha=1/2$, $S_1(\tau)=\tau/2+\sin(2\tau)/4$,
and Eq. \eqref{eq:model:fconst:LC} becomes
\begin{equation}
{Y}_{1} (\tau)  =
\frac{e^{\pi\sigma/4}}{e^{\pi\sigma/2}+1}
\exp\left(-\frac{\sigma}{4}\sin(2\tau)\right)
\!\!\!
\int\limits_{-\pi/2}^{\pi/2}
\!\!\!\!
\exp\left(\frac{\sigma}{2} \eta\right)
\exp\left(-\frac{\sigma}{4}\sin(2\eta+2\tau)\right)
\sin(\eta \! + \! \tau)d\eta
.
\label{eq:Y1}
\end{equation}
Using the generating function for the modified Bessel function of the first kind \cite{math},
\begin{equation}
\exp\left[ \frac{z}{2} \left(p+\frac{1}{p}\right) \right]
=
\sum\limits_{k=-\infty}^{\infty} I_k(z) p^k
,
\end{equation}
we change exponent terms involving $\sin$ function into series
in the following way
\begin{equation}
\exp\left(-\frac{\sigma}{4}\sin(2\zeta)\right)
=
\sum\limits_{k=-\infty}^{\infty} I_k\left(\frac{\sigma}{4}\right) i^k e^{2ik\zeta}
.
\end{equation}
Then the right-hand side of Eq. \eqref{eq:Y1} is transformed into a double series
\begin{equation}
{Y}_{1} (\tau)  =
\frac{e^{\pi\sigma/4}}{e^{\pi\sigma/2}+1}
\sum\limits_{m=-\infty}^{\infty} 
\sum\limits_{k=-\infty}^{\infty} 
i^{m+k}
I_m\left(\frac{\sigma}{4}\right) 
I_k\left(\frac{\sigma}{4}\right)
e^{2im\tau}
\!\!\!
\int\limits_{-\pi/2}^{\pi/2}
\!\!\!\!
e^{\sigma\eta/2 + 2ik(\eta+\tau)}
\sin(\eta \! + \! \tau)d\eta
.
\label{eq:Y1ds}
\end{equation}
Integrating term-by-term and rearranging the sums
by collecting terms involving the same harmonics,
it is not difficult to obtain the following formulae
\begin{gather}
Y_1(\tau) = \sum\limits_{m=-\infty}^{\infty} \left(
C_m e^{i (2m-1) \tau} + C_{1-m} e^{-i (2m-1) \tau}
\right),
\label{eq:Y1s}
\\
C_m = i^m
\sum\limits_{k=-\infty}^{\infty}
\frac{(-1)^{k+1}}{4k+2-i\sigma}
\left[
I_k\left(\frac{\sigma}{4}\right) - i I_{k+1}\left(\frac{\sigma}{4}\right)
\right] I_{k+1-m}\left(\frac{\sigma}{4}\right).
\label{eq:Cms}
\end{gather}
Using Eq. \eqref{eq:Cms} and
the symmetry property $I_{-k}(z)=I_k(z)$ \cite{math},
it is easy to show that $C_{1-m}=C^*_m$.
For $m=1$ we obtain Eq. \eqref{eq:C1} from the main text:
\begin{equation}
C_1=
\!\!\!
\sum\limits_{k=-\infty}^{\infty}
\!\!\!
\frac{(-1)^{k+1}}{4k+2-i\sigma}
\left[
i I_k^2\left(\frac{\sigma}{4}\right) + I_{k}\left(\frac{\sigma}{4}\right)I_{k+1}\left(\frac{\sigma}{4}\right)
\right].
\tag{{25}}
\end{equation}
The representation Eq. \eqref{eq:Cms} is equivalent to
Eq. \eqref{eq:Cm} provided that $C_1$ is defined by Eq. \eqref{eq:C1}.

\begin{figure}
\includegraphics[width=0.6\columnwidth]{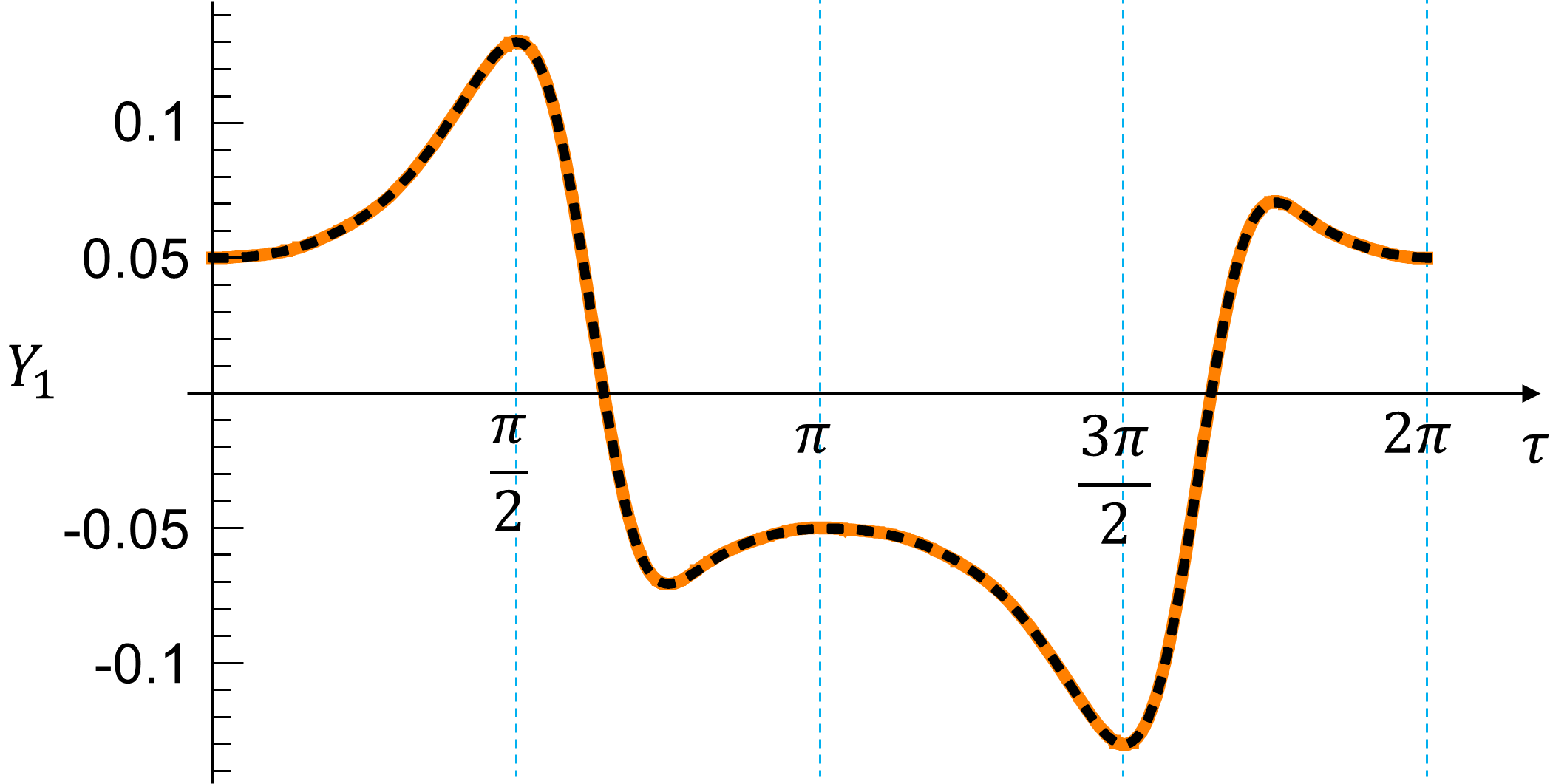}
\caption{Solution of  Eq. \eqref{eq:n2} for $\sigma=20$.
Solid orange curve: analytical solution
given by  Eqs. \eqref{eq:LCF}, \eqref{eq:C1}, and \eqref{eq:Cm}.
Black dashed curve: numerical solution.
\label{fig:1}}
\end{figure}

Fig. \ref{fig:1} shows a comparison of the numerical solution of Eq. \eqref{eq:n2}
for $\sigma=20$ with the analytical solution defined
by Eqs. \eqref{eq:LCF}, \eqref{eq:C1}, and \eqref{eq:Cm},
where the sum in Eq. \eqref{eq:LCF} is cut at $m=6$
(which corresponds to the $2m-1=11^{\rm th}$ harmonic)
and the sum in Eq. \eqref{eq:C1} is cut at $k=\pm 10$
(i.e. only terms with the index $k$ from -10 to 10 are taken into account).


\end{document}